\documentclass{article}
\usepackage{fullpage}
\usepackage{amsmath,amssymb}
\usepackage{amsthm}
\usepackage{subfigure}
\usepackage{xcolor}
\usepackage{verbatim}
\usepackage{pst-all}
\usepackage{cite}

\newcommand{\coopGame}{\Delta}
\newcommand{\game}{\Gamma}
\newcommand{\real}{\mathbb{R}}
\newcommand{\nat}{\mathbb{N}}

\newcommand{\D}{\displaystyle}

\newcommand{\classNP}{{\sc NP}}
\newcommand{\classP}{{\sc P}}

\newtheorem{theorem}{Theorem}
\newtheorem{lemma}{Lemma}
\newtheorem{corollary}{Corollary}
\newtheorem{proposition}{Proposition}

\begin{document}
\title{Strategic Cooperation in Cost Sharing Games} \author{Martin
  Hoefer\thanks{Dept. of Computer Science, RWTH Aachen University,
    Germany, {\tt mhoefer@cs.rwth-aachen.de}. Supported by DFG through
    UMIC Research Centre at RWTH Aachen University.}}
\date{}
\maketitle

\begin{abstract}
  In this paper we consider strategic cost sharing games with
  so-called \emph{arbitrary sharing} based on various combinatorial
  optimization problems, such as vertex and set cover, facility
  location, and network design problems. We concentrate on the
  existence and computational complexity of strong equilibria, in
  which no coalition can improve the cost of each of its members.

  Our main result reveals a connection between strong equilibrium in
  strategic games and the core in traditional coalitional cost sharing
  games studied in economics. For set cover and facility location
  games this results in a tight characterization of the existence of
  strong equilibrium using the integrality gap of suitable linear
  programming formulations. Furthermore, it allows to derive all
  existing results for strong equilibria in network design cost
  sharing games with arbitrary sharing via a unified approach. In
  addition, we are able to show that in general the strong price of
  anarchy is always 1. This should be contrasted with the price of
  anarchy of $\Theta(n)$ for Nash equilibria. Finally, we indicate
  that the LP-approach can also be used to compute near-optimal and
  near-stable approximate strong equilibria.
\end{abstract}

\section{Introduction}

How can a set of self-interested actors share the cost of a joint
investment in a fair and stable way? This fundamental question has
motivated a large amount of research in economics in the last
decades~\cite{Young94}. More recently, this question has been studied
in computer science to understand the development of the Internet and
questions arising in e-commerce~\cite{JainChapter07}. A classic
framework to study cost sharing problems without centralized control
are \emph{cost sharing games}, in which cost can be specified as an
abstract parameter for each player and/or each coalition. Relevant to
real-world optimization problems are cost sharing games, where the
cost is tied to the investment into specific resources. Such games
based on combinatorial optimization problems have a long tradition in
economics. They are usually formulated as coalitional game, i.e.,
there is a set of players, and each coalition of players has an
associated cost value. This value comes from an optimal solution to an
underlying optimization problem for the coalition. For example,
consider a multicast network design game, in which players strive to
establish connections to a common source vertex $s$. This scenario can
be formulated as a MST game, in which each vertex $v \neq s$ is a
player, and the edges have costs. The cost of a coalition in this game
is the cost of the MST for the set $C \cup s$. In the literature many
interesting and important coalitional cost sharing games have been
studied, e.g., based on problems like MST and Steiner
tree~\cite{Bird76,Megiddo78,Granot81,Tamir91,Granot98,Faigle97},
covering and packing problems~\cite{Deng99}, facility
location~\cite{Tamir93,GoeSku04}, or TSP~\cite{Faigle98}.
The problem in these games is to devise a stable and fair way of
sharing the cost among the players. Coalitional cost sharing games are
usually \emph{transferable utility (TU) games}, i.e., the cost can be
shared arbitrarily between the players. This allows for the largest
level of generality for possible interactions in the bargaining and
coalition formation process. The foremost concept of stability and
fairness in TU cost sharing games is the \emph{core}. The core is a
set of imputations, i.e., of distributions of the cost for the
complete player set to the players. To be in the core an imputation
has to fulfill the additional property that no coalition of players in
sum pays more than its associated cost value. Results about the
non-emptiness of the core and characterizations of core solutions have
been obtained for many of the games mentioned above.
%
%

A problem with coalitional cost sharing games is that cost shares
represent a strong abstraction from the underlying optimization
problem. Players are assumed to contribute on a global level, and the
game does not take into account who pays how much for which
resource. Such information, however, is crucial when studying the
incentives of players in large unregulated settings such as the
Internet. The need to understand cost sharing on a strategic level
prompted computer scientists to study strategic cost sharing
scenarios. On the one hand, there are a number of recent works on
\emph{designing} strategic cost sharing games to obtain favorable Nash
equilibrium properties~\cite{ChenRough10}. In these games the
underlying assumption is that a central authority designs and
maintains the solution and dictates cost shares for each player. This
is close to assumptions that are made in the area of cost sharing
mechanisms, which have received a lot of
attention~\cite{JaVa01,Pal03,Immorlica05,Konemann08}. Designing cost
shares, e.g., using Shapley value cost
sharing~\cite{Anshe08,Albers09,Leonardi07,Epstein09}, can yield
favorable properties on the existence and cost of Nash equilibria. In
contrast, such a model is unsuitable when there is very little control
over players and their bargaining options. A model that allows for
general cost sharing between players is sometimes referred to as
\emph{arbitrary} cost sharing, and it has been studied
in~\cite{Anshe03, Anshe09, HoeferISAAC06, CarHoe06, Hoefer09,
  Hoefer08, Hoefer05, Epstein09,AnsheCas09}. In these cost sharing games
the strategy of a player is a payment function that specifies his
exact contribution to the cost of each resource. The outcomes of such
strategic cost sharing games based on combinatorial optimization
problems will be the subject of this paper.

When studying the outcomes of the interaction of rational agents in
strategic games, we need to discuss the appropriate solution
concept. The most prominent stability concept in strategic games is
the Nash equilibrium (NE). While a NE (in mixed strategies) always
exists, a drawback is that it is only resilient to unilateral
deviations. In many reasonable scenarios agents might be able to
coordinate their actions, and under these circumstances a NE is not a
reasonable solution concept. To address this issue we consider the
\emph{strong equilibrium (SE)} in this paper. A strong
equilibrium~\cite{Aumann59} is a state, from which no coalition (of
arbitrary size) has a deviation that lowers the cost of \emph{every}
member of the coalition. This resilience to coalitional deviations is
highly attractive. On the downside, strong equilibria might not exist
in a game. This may be the reason they have not received an equivalent
amount of research interest despite their attractive properties. We
partly circumvent this problem by studying approximate versions of the
strong equilibrium, which is guaranteed to exist. However, our
treatment of these aspects is brief and mostly left for future work.

Our main interest is to characterize the existence, social cost, and
computational complexity of SE in strategic cost sharing games based
on combinatorial optimization problems. Our first insight in
Section~\ref{sec:PoA} reveals that the concept of SE in strategic cost
sharing games is equivalent to seemingly stronger notions of
super-strong or sum-strong equilibria. Additionally, we show that a SE
in a strategic game can always be turned into a core imputation of the
corresponding coalitional game defined on the same instance of the
optimization problem. Hence, a SE represents a \emph{strategic
  refinement} of a core solution, and existence of a SE implies
non-emptiness of the core. It also implies that the strong price of
anarchy~\cite{Andelman09} is 1, i.e., in every SE a solution is bought
that is a social optimum to the underyling optimization problem.

In Section~\ref{sec:LP} we consider a variety of games based on vertex
and set cover and various facility location problems. For these games
we show an equivalence result of core and SE. In particular, whenever
the core in the coalitional game is nonempty, there is a SE for the
strategic game. Our main proof technique relies on a connection via
linear programming to Owen's linear production model~\cite{Owen75},
which is one of the most common ways to show non-emptiness of the core
in coalitional games~\cite{Deng99}. Using this machinery we are able
to tightly characterize the existence and cost of SE in all these
games. Our results extend to special classes of network design
problems. This includes, e.g., MST and classes of Steiner Network
Design Games~\cite{Anshe03,Hoefer09}. As a byproduct our general
approach yields simple proofs for all known results for SE in
strategic cost sharing games with arbitrary sharing, which were
previously shown~\cite{Epstein09} via complicated combinatorial
arguments.

The equivalence between SE and core solutions is an interesting and
notable fact. However, in other cases such as cooperative games with
non-transferable utility (NTU games) and appropriate extensions to
strategic games a similar equivalence is obvious. Thus, it may be more
surprising to observe that the relation between SE and core solutions
in cost sharing games can be quite complicated. In particular, in
Section~\ref{sec:Comb} we explore equivalence without relying on
linear programming. While in some cases like Terminal Backup
Games~\cite{Anshe09} we can resort to combinatorial arguments, in
other interesting games our results are mostly negative. In
particular, we show that in Steiner Tree connection games or network
cutting games equivalence does not hold, i.e., the core might be
non-empty but a SE is absent. A similar result is established in
Section~\ref{sec:Fractional} even for simple vertex cover games when
we allow resources to be purchased fractionally or in mulitple
units. Characterizing SE in these games remains as an intriguing open
problem. We observe in Section~\ref{sec:Approx} that linear
programming can be used to obtain approximate $(\alpha,\beta)$-SE in
vertex and set cover, as well as factility location games. Finally, we
conclude in Section~\ref{sec:Conclude} with some interesting questions
for further research.

Our main conceptual contribution is to reveal a non-trivial and close
relation between coalitional and strategic games defined on the same
instance of the optimization problem. The strategic game can be seen
as a strategic variant of the coalitional game. In particular, the
cost value of a coalition is the minimax cost level that the coalition
can achieve in the strategic game. In addition, our analysis reveals
that in many games SE can act as a strategic refinement of rather
coarse core solutions. We believe that this inherent connection
between traditional coalitional games studied in economics and
strategic cost sharing games recently formulated in computer science
is of independent interest and should stimulate further research on
cost sharing with rational agents.

\subsection{Preliminaries}

We consider classes of cost sharing games based on combinatorial
optimization problems. In each of these games there is a set $R$ of
resources. Resource $r \in R$ can be \emph{bought} if the associated
cost $c(r) \ge 0$ is paid for. For $R' \subseteq R$ let $c(R') =
\sum_{r \in R'} c(r)$. We assume that there is set of players
$K$. Each player $k \in K$ strives to satisfy a certain constraint on
the bought resources. For example, in the case of the \emph{set cover
  problem} the player set is the element set $K = E$. The resources
are sets $R = \mathcal{S} \subseteq 2^E$ over $E$. The constraint of
player $e$ states that there must be at least one bought set $S$ with
$e \in S$. In a similar way we can base our construction on various
cost minimization problems like facility location or network
design. We will describe them in more detail in the corresponding
sections. However, a common assumption in our problems is a free
disposal property, i.e., if for a set of bought resources all player
constraints are satisfied, then a superset of bought resources can
never make a player constraint become violated.

For a given set of players, resources, and constraints we define two
games - a \emph{coalitional} and a \emph{strategic} cost sharing
game. The \emph{coalitional game} $\coopGame = (K,c)$ is given by the
set of players $K$ and a cost function $c : 2^K \to \real_0^+$ that
specifies a cost value for every subset of players. For a coalition $C
\subseteq K$, the cost is $c(C) = \sum_{r \in R(C)^*} c(r)$ for an
\emph{optimum solution} $R(C)^* \subseteq R$ for $C$. In particular,
$R(C)^*$ is a minimum cost set of resources that must be bought to
satisfy all constraints of players in $C$. For example, in a set cover
game $R(C)^*$ is the minimum cost set cover for the elements in
$C$. We denote the special case $R^* = R(K)^*$ as the \emph{social
  optimum}.

The goal in a coalitional game is to find a cost sharing of $c(K)$ for
the so-called grand coalition $K$. A vector of cost shares
$\gamma_1,\ldots,\gamma_k$ is called an \emph{imputation} if $\sum_{i
  \in K} \gamma_i = c(K)$. The game $\coopGame$ is a
\emph{transferable utility (TU)} game, i.e., we are free to choose $0
\le \gamma_i \le c(K)$. The central concept of stability and fairness
in coalitional games is the \emph{core}. The core is the set of
imputations $\gamma$, for which $c(C) \ge \sum_{i \in C}
\gamma_i$. Intuitively, when sharing the cost according to a member of
the core, no subset of players has an incentive to deviate from the
grand coalition and make a separate investment - depending on the
underlying optimization problem, e.g., purchase different sets or
construct an independent network.

The \emph{strategic game} $\game = (K, (S_i)_{i \in K}, (c_i)_{i \in
  K})$ is specified by strategies and individual cost for each
player. The \emph{strategy space} $S_i$ of player $i \in K$ consists
of all functions $s_i : R \to \real_0^+$. Strategy $s_i$ allows him to
specify for each resource $r \in R$ how much he is willing to
contribute to $r$. A resource $r$ is \emph{bought} if $\sum_{i \in K}
s_i(r) \ge c(r)$. A vector of strategies $s$ is a \emph{state} of the
game. For a state $s$ we define $|s_i| = \sum_{r \in R} s_i(r)$ and
the \emph{individual cost} of player $i$ as $c_i(s) = |s_i|$ if the
bought resources satisfy his constraint. Otherwise, $c_i(s) = \infty$
or a different value that is prohibitively large.

The foremost concept of stability in strategic games is the Nash
equilibrium, a state in which no player unilaterally has an incentive
to deviate. In this paper, however, we consider coalitional incentives
and thus resort to a strengthened version called strong
equilibrium~\cite{Aumann59}. A state $s$ has a \emph{violating
  coalition} $C \subseteq K$ if there are strategies $s'_C = (s'_i)_{i
  \in C}$ such that $c_i(s'_C,s_{-C}) < c_i(s)$ for each $i \in C$. A
violating coalition has a deviation, in which all players in $C$
strictly pay less. A \emph{strong equilibrium} is a state $s$ that has
no violating coalition. Note that in a SE a set of resources is bought
such that all player constraints are satisfied. Each resource $r$ is
either paid for exactly or not contributed to at all. Thus, a SE
represents a cost sharing of a feasible solution for the grand
coalition. In addition, we briefly consider the concept of a
$(\alpha,\beta)$-SE. These are strategy profiles, which constitute an
approximate solution concept. In a $(\alpha,\beta)$-SE no coalition of
players can reduce the cost of every member by strictly more than a
factor of $\alpha$, and the cost of the bought solution represents a
$\beta$-approximation to $c(K)$.

\section{Strong Equilibria and the Core}
\label{sec:PoA}

Consider a given set of resources $R$ with costs $c(r)$ and a set of
players $K$ with constraints. Our first result reveals that several
coalitional equilibrium concepts coincide in strategic games
$\game$. In particular, we consider \emph{super-strong} and
\emph{sum-strong equilibria}. A state $s$ has a \emph{weakly violating
  coalition} $C \subseteq K$ if there are strategies $s'_C = (s'_i)_{i
  \in C}$ such that $c_i(s'_C,s_{-C}) \le c_i(s)$ for each $i \in C$
and $c_{i'}(s',s_{-C}) < c_{i'}(s)$ for at least one $i' \in C$. A
state $s$ has a \emph{sum violating coalition} $C \subseteq K$ if
there are strategies $s'_C = (s'_i)_{i \in C}$ such that $\sum_{i \in
  C} c_i(s'_C,s_{-C}) < \sum_{i \in C} c_i(s)$. A super-strong
(sum-strong) equilibrium is a state $s$ that has no weakly (sum)
violating coalition. Note that every violating coalition is also
weakly violating, and every weakly violating coalition is also sum
violating. Hence, every sum-strong equilibrium is super-strong, and
every super-strong equilibrium is strong. We note this simple fact
because we actually show the absence of sum violating coalitions in
our proofs below. In general strategic games it is easy to see that
the inclusions are strict, i.e., a strong equilibrium might not be
sum-strong. In our strategic cost sharing games $\game$, however,
every strong equilibrium is sum-strong.

\begin{proposition}
  \label{prop:sumstrong}
  Every strong equilibrium in a strategic game $\game$ is a
  sum-strong equilibrium.
\end{proposition}

\begin{proof}
  We will show that if a state has a sum violating coalition, it also
  has a violating coalition. Suppose in a state $s$ there is a sum
  violating coalition $C$ and a corresponding deviation $s'_C$. Note
  that this directly implies $\sum_{i \in C} c_i(s) > 0$ and
  $c_i(s'_C,s_{-C}) = |s'_i|$ for all $i \in C$. If a player $i \in C$
  has prohibitively large cost in $s$, he has a unilateral deviation
  purchasing all resources in $R^*$ by himself, which obviously yields
  a violating (singleton) coalition. Hence, we also assume that for
  all players $i \in C$ we have finite cost $c_i(s) = |s_i|$ in $s$.

  First consider players $i \in C$, for which $|s_i| = 0$, and denote
  the set of these players by $C_0$. We can construct a state with
  $|s''_i| = 0$ for $i \in C_0$ as follows. We define $s'(r) = \sum_{i
    \in C} s'_i(r)$ and let each player pay $s''_i(r) = (|s_i|/\sum_{i
    \in C} |s_i|) \cdot s'(r)$. This yields $|s''_i| = 0$ if and only
  if $|s_i| = 0$.

  Now we define the coalition $C_1 = C - C_0$ and the deviation
  $s''_{C_1} = (s''_i)_{i \in C_1}$. For every player $i \in C_1$ we
  have
  \[ |s''_i| = \sum_{r \in R} \frac{|s_i|}{\sum_{i \in C} |s_i|} \cdot
  s'(r) = |s_i| \cdot \frac{\sum_{i \in C_0} |s'_i|}{\sum_{i \in C}
    |s_k|} = |s_i| \cdot \frac{\sum_{i \in C} c_i(s'_C,
    s_{-C})}{\sum_{i \in C} c_i(s)} < |s_i|\enspace,
  \]
  using the assumption for $s'_C$. Note that for each resource $r$ we
  have same total contribution in $s'_C$ and $s''_{C_1}$, hence for
  every $i \in C_1$ we have $c_i(s''_{C_1}, s_{-C_1}) = |s''_i| <
  |s_i| = c_i(s)$. This implies that $C_1$ is a violating coalition.
\end{proof}

We note on the side that for every state $s$, coalition $C$, deviation
$s'_C = (s'_i)_{i \in C}$, and finite $\alpha \ge 1$ with $\sum_{i \in
  C} c_i(s'_C,s_{-C}) = \alpha \cdot \sum_{i \in C} c_i(s)$ we can
find in a similar way $C'$ and $s''_{C'} = (s''_i)_{i \in C'}$ with
$c_i(s''_C,s''_{-C}) = \alpha c_i(s)$ for \emph{every} $i \in
C'$. Thus, the equivalence of strong and sum-strong equilibria holds
also for approximate versions of the concepts, in which players must
improve their costs by a factor of strictly more than $\alpha$.

We continue to show a general connection between core imputations for
the coalitional game $\coopGame$ and SE of the strategic game
$\game$. We first observe that in a SE players always share the cost
of a social optimum $R^*$.

\begin{theorem}
  \label{theo:SPoA}
  In every strong equilibrium of a strategic game $\game$ the players
  share the cost of a social optimum. The strong price of anarchy is
  1.
\end{theorem}

\begin{proof}
  Consider a SE $s$ and the set $R'$ of bought resources. Assume for
  contradiction $c(R') > c(R^*)$. If all players jointly deviate to
  purchase $R^*$, each player must pay only a fraction of
  $c(R^*)/c(R') < 1$ of $|s_k|$. A player $k$ chooses to pay $s'_k(r)
  = c(r) \cdot \frac{|s_k|}{c(R')}$. If all players jointly deviate to
  $s'$, this obviously strictly decreases the payment of \emph{all}
  players. Hence, if $c(R') > c(R^*)$, then $K$ is a violating
  coalition for $s$, a contradiction.
\end{proof}

\begin{proposition}
  \label{prop:SEisCore}
  If the strategic game $\game$ has a strong equilibrium, then
  the coalitional game $\coopGame$ has a core solution.
\end{proposition}

\begin{proof}
  Consider a SE $s$ of $\game$, which by Theorem~\ref{theo:SPoA}
  is a cost sharing of $R^*$, and a coalition $C$. The coalition has
  the possibility to deviate and contribute just to buy $R(C)^*$. In
  this case it has to share for each resource $r \in R(C)^*$ at most
  the remaining cost on top of the contribution of players in $K
  \backslash C$, i.e., $c_C(r) = c(r) - \sum_{k \in K \backslash C}
  s_k(r)$. If $c_C(R(C)^*) < \sum_{k \in C} |s_k|$, the coalition can
  deviate to $s'_k(r) = c_{C}(r)\cdot\frac{|s_k|}{c(R^*)}$ for every
  $k \in C$ and every $r \in R(C)^*$, which would represent an
  improvement for every player in $C$. However, as $s$ is a SE, $C$
  must not be violating, and so $c_C(R(C)^*) \ge \sum_{k \in C}
  |s_k|$. Trivially, $c(r) \ge c_C(r)$, and so $c(R(C)^*) \ge \sum_{k
    \in C} |s_k|$. Thus, $\gamma$ with $\gamma_k = |s_k|$ is in the
  core of $\coopGame$.
\end{proof}

These two simple insights show that non-emptiness of the core is a
necessary condition for existence of a SE. In the following we
consider various classes of games, in which it is also sufficient. In
these cases the SE is a strategic refinement of the core, as it allows
to specify a strategic allocation of payments to resources.

\section{Strong Equilibria using Linear Programming}
\label{sec:LP}
\subsection{Vertex and Set Cover Games}

In a variety of fundmental games non-emptiness of the core and
existence of SE are equivalent. We can relate SE existence to the core
via linear programming duality. For simplicity we outline the general
argument in the setting of set cover games. In a set cover game, we
are given a set of players as elements $E$ and a set system
$\mathcal{S} \subseteq 2^E$, where each $S \in \mathcal{S}$ has a cost
$c(S) \ge 0$. The constraint of player $e$ is that at least one set
$S$ with $e \in S$ must be bought.

\begin{theorem}
  If a set cover game $\coopGame$ has a non-empty core, then the
  strategic game $\game$ has a strong equilibrium.
\end{theorem}

\begin{proof}
  We consider the integer programming formulation of set cover. In
  particular, we consider the following linear relaxation, which
  employs $x_S \ge 0$ instead of $x_S \in \{0,1\}$ and thus allows
  sets to be included fractionally in the solution. 

  \begin{equation*}
    \begin{array}{llr}
      \mbox{Min } & \D \sum_{S \in \mathcal{S}} x_S c(S) & \\
      \mbox{subject to } & \D\sum_{S : e \in S} x_S \ge 1 & \forall \; e \in E  \\
      & x_S \ge 0 & \forall \; S \in \mathcal{S}.
    \end{array}
  \end{equation*}
  We also consider the corresponding LP dual.
  \begin{equation*}
    \begin{array}{llr}
      \mbox{Max } & \D \sum_{e \in E} \gamma_e & \\
      \mbox{subject to } & \D \sum_{e \in S} \gamma_e \le c(S) & \forall \; S \in \mathcal{S}\\
      & \gamma_e \ge 0 & \forall \; e \in E.
    \end{array}
  \end{equation*}
  It has been shown by Deng et al.~\cite{Deng99} that the core of
  $\coopGame$ is non-empty if and only if the integrality gap of the
  underlying set cover problem is 1, i.e., if the LP has an integral
  optimal solution. With Proposition~\ref{prop:SEisCore} this is a
  prerequisite for existence of a SE in $\game$. We strengthen this
  result by showing that core solutions can also be turned into an
  allocation of payments to resources for a SE in $\game$. Thus, an
  integral optimum is also sufficient.

  For the above programs consider the optimum primal solution $x^*$
  and the optimum dual solution $\gamma^*$, where $x^*$ is integral
  and defines a feasible cover. Both $x^*$ for the primal and
  $\gamma^*$ for the dual yield the same objective value. Now assign
  each player $e$ to pay $s_e(S) = \gamma_e^* x_S^*$. The theorem
  follows if every set in the cover is purchased exactly and no
  coalition $C$ can reduce their total payments $\sum_{e \in C}
  |s_e|$. The first condition is clearly necessary for a SE, the
  second one implies that no coalition can be sum violating (and thus
  violating). We first show that the sets are exactly paid for. If
  $x_S^* > 0$, then due to complementary slackness the inequality
  $\sum_{e \in E} \gamma_e^* \le c(S)$ is tight, hence by this
  assignment all the purchased sets get exactly paid for.

  We now show that no coaltion can reduce the total payments. Suppose
  a coalition $C$ is violating. We consider an adjusted game derived
  by iteratively removing elements and payments of other players $e
  \not\in C$. Upon removing an element $e$, we remove its contribution
  from the costs of sets $S$ including $e$. This yields the cost
  function $c_C(S)$ with
  \[
  c_C(S) = c(S) - \sum_{e \not\in C, e \in S} \gamma_e^*x_S^*\enspace.
  \]
  It captures the reduced problem of finding a minimum cost cover for
  coalition $C$ with costs adjuted by the payments of other players $e
  \not\in C$. Note that for this reduced problem the solution $x^*$ is
  still feasible. By obtaining the dual we can set the covering
  requirement to $0$ for every removed element $e \not\in C$. Then
  $\gamma^*$ still represents a feasible solution to the LP-dual of
  the reduced problem. It yields the same objective value as $x^*$ for
  the primal. By strong duality both $x^*$ and $\gamma^*$ must be
  optimal solutions to the reduced primal and dual problems. This
  proves that the total payments of $C$ are optimal. Hence, $C$ cannot
  be sum violating and not violating, a contradiction. This proves
  that $s$ is a SE.
\end{proof}

The main idea of the proof is to use duality arguments for a cost
reduction of resources. In particular, for an optimum $x^*$, the
objective function can be represented by a linear combination of tight
constraints. The multipliers are the optimal dual variables
$\gamma^*$. This can be used to reduce the costs and show optimality
of $x^*$ even for every coalition.

For the special case of vertex cover games we can use results
from~\cite{Deng99} to efficiently compute SE. In particular, a game
allows a core solution (and thus a SE) if and only if a maximum
matching in the graph has the same size as the minimum vertex
cover. This condition can be checked in polynomial time by computing
corresponding vertex covers and matchings~\cite[Theorem 7 and
Corollary 7]{Deng99}. Hence, we can check in polynomial time whether a
SE exists. If it exists, we can use the computed vertex cover as
primal solution for our LP and compute cost shares for a strong
equilibrium with the corresponding dual solution.
\begin{corollary}
  In a vertex cover game we can decide in polynomial time if a strong
  equilibrium exists. If it exists, we can compute a strong
  equilibrium in polynomial time.
\end{corollary}
In addition, we can check in polynomial time whether a given strategy
profile is a SE.
\begin{corollary}
  Given a state $s$ for a vertex cover game $\game$ we can verify in
  polynomial time if it is a strong equilibrium.
\end{corollary}
If the strategy profile is a SE, it must exactly pay for a vertex
cover of the problem. This yields a primal solution for the LP. In
addition, the accumulated cost shares of players must yield a
corresponding dual solution. Finally, both primal and dual solutions
must generate the same value of the objective function. This is a
sufficient and necessary condition for being a SE, which can be
checked in polynomial time.

Another interesting case are edge cover games. Here players are the
vertices of a graph and resources are the edges. Each vertex wants to
ensure that at least one incident edge is bought. Using the
characterization of the non-emptiness of the core in~\cite[Theorem 8
and Corollary 8]{Deng99} we can obtain similar results for this game
as well.

\begin{corollary}
  In an edge cover game $\game$ we can decide in polynomial time if a
  strong equilibrium exists. If it exists, we can compute a strong
  equilibrium in polynomial time. Given a state $s$ for an edge cover
  game $\game$ we can verify in polynomial time if it is a strong
  equilibrium.
\end{corollary}

\subsection{Facility Location Games}

Another class of games that can be handled via similar arguments are
facility location games. We outline the arguments on the simple class
of \emph{uncapacitated facility location games} (UFL games) and show
below how to extend this approach to a more general class of games
considered in~\cite{GoeSku04,HoeferISAAC06}. In a \emph{UFL problem}
there is a set $T$ of terminals and a set $F$ of facilities. We set
$n_t = |T|$ and $n_f = |F|$. Each facility $f \in F$ has an opening
cost $c(f) \ge 0$, for each terminal $t \in T$ and each facility $f
\in F$ there is a connection cost $c(t,f) \ge 0$. The goal is to open
a subset of facilities and buy a set of connections of minimum total
cost, such that each terminal is connected to an opened facility. In
the \emph{UFL game} each player owns a terminal, i.e., $K = T$. The
constraint of player $t$ is satisfied if there is a bought connection
$(t,f)$ to some opened facility $f$. We can formalize the UFL problem
by an integer program as follows:
\begin{equation*}
  \begin{array}{lll}
    \mbox{Min } & \multicolumn{2}{l}{\D \sum_{f \in F} c(f)y_f + \sum_{t \in T} c(t,f)x_{tf}} \\
    \mbox{subject to } & \D \sum_{f \in F} x_{tf} \ge 1 & \mbox{ for all }
    t \in T\\
    & y_f - x_{tf} \ge 0 & \mbox{ for all } t \in T, f \in F\\  
    & y_f, x_{tf} \in \{0,1\} & \mbox{ for all } t \in T, f \in F, 
  \end{array}
\end{equation*}
\begin{theorem}
  If a UFL game $\coopGame$ has a non-empty core, then the strategic
  game $\game$ has a strong equilibrium.
\end{theorem}

\begin{proof}
  We again use the linear relaxation, which can be obtained by
  replacing $y_f, x_{tf} \in \{0,1\}$ by $y_f, x_{tf} \ge 0$. Then the
  dual can be given by
  \begin{equation*}
    \begin{array}{lll}
      \mbox{Max } & \multicolumn{2}{l}{\D \sum_{t \in T} \gamma_t} \\
      \mbox{subject to } & \D \sum_{t \in T} \gamma_t - \delta_{tf} \le c(t,f) &
      \mbox{ for all } t \in T, f \in F \\
      & \sum_{t \in T} \delta_{tf} \le c(f) & \mbox{ for all } f \in F.
    \end{array}
  \end{equation*}
  It has been shown by Goemans and Skutella~\cite{GoeSku04} that the
  core of $\coopGame$ is non-empty if and only if the integrality gap
  of this LP is 1. We can now argue similarly as before. An integral
  optimum solution $(x^*, y^*)$ to the LP-relaxation represents a
  partition of the terminal set $T$ into a collection of stars, one
  for each facility $f$. The constraints corresponding to these sets
  hold with tightness, and we can assign each player $t$ to pay for
  her terminal the amount $s_t(t,f) = (\gamma_t^* -
  \delta_{tf}^*)x_{tf}^*$ as connection cost to $f$, in which
  $(\gamma^*, \delta^*)$ is the optimum solution to the dual. For the
  opening costs $s_t(f) = \delta_{tf}^*y_f^*$. In total this pays
  exactly for all costs of the solution by duality.

  Suppose there is a violating coalition $C$. We again remove players
  in $K - C$ and reduce the costs of connections and facilities by the
  respective contributions. In order to represent a violating
  coalition, the players in $C$ must be able to deviate and reduce
  their total sum of payments. However, the solution $(x^*,y^*)$ has
  the same value for the reduced LP of coalition $C$ as
  $(\gamma^*,\delta^*)$ for the dual of the reduced LP. By duality
  both solutions remain optimal. Thus, coalition $C$ is purchasing an
  optimal solution against the payments of players in $K - C$ and has
  no possibility to reduce the total payments. This is a contradiction
  to $C$ being a violating coalition.
\end{proof}

This result can be combined with insights from~\cite{GoeSku04} to
characterize computational properties of SE. In particular, we can
decide in polynomial time if a given strategy profile for $\game$ is a
SE. We first check if the payments of players are made only to their
own connection and opening costs. Then we accumulate contributions to
cost shares and check if this yields a core solution - i.e., if the
primal solution (given by the purchased solution to the facility
location problem) and the dual solution (given by the cost shares)
correspond to each other and yield the same optimal value for primal
and dual LPs.

\begin{corollary}
  Given a strategy vector for a UFL game $\game$ we can verify in
  polynomial time if it is a strong equilibrium.
\end{corollary}

As verfication is in \classP, the problem of computing a strong
equilibrium is in \classNP. In fact, in~\cite{GoeSku04} it is shown
for a class of UFL games that deciding the existence of a core
solution is \classNP-complete. As existence of SE and core solutions
is equivalent, this yields the following result.

\begin{corollary}
  It is \classNP-complete to decide if a given UFL game $\game$ has a
  strong equilibrium.
\end{corollary}

In the Appendix we show how to extend this result to
connection-restricted facility location games (CRFL games)
from~\cite{GoeSku04}, in which access to a facility $f$ can be
obtained only by certain allowed coalitions $\mathcal{A}_f \subseteq
2^T$. We consider the special case of \emph{closed} games (CCRFL
games), in which the set system $\mathcal{A}_f$ of allowed coalitions
is downward closed, i.e., each subset of an allowed coalition is also
an allowed coalition. This simplifies the specific allocation of the
cost shares to connections and facilities. While the closed property
is a restriction, we note that many variants of facility location
arising in practice fall into this class of games, e.g., problems with
capacity or incompatibility constraints. We believe that equivalence
between core and SE also holds for CRFL games in full generality, but
a proof of this statement remains as an open problem.

\begin{theorem}
  If a CCRFL game $\coopGame$ has a non-empty core, then the strategic
  game $\game$ has a strong equilibrium. Given a strategy vector for a
  CCRFL game $\game$ we can verify in polynomial time if it is a
  strong equilibrium.
\end{theorem}

\subsection{Connection Games}

In this section we use a linear program to formulate network design
games in directed and undirected graphs. Perhaps the most frequently
studied variant is a \emph{connection game} originally formulated
in~\cite{Anshe03}. In this game there is a graph $G=(V,E)$, resources
are the edges, and each edge has a non-negative cost $c(e) \ge
0$. There is a set of players $K$, and each player $k$ has a
source-sink pair $(s_k, t_k)$. A player is satisfied if there is a
path of bought edges connecting his source and sink. This is a game
based on the Steiner network problem in graphs~\cite{GoeWil95}. In a
variant based on Steiner Tree called the \emph{single source} game,
every player has the same source $s$. Here we characterize existence
of SE based on a Flow-LP previously studied, e.g.,
in~\cite{Wong84,Tamir91}.
\begin{theorem}
  If the Flow-LP has an integral optimum solution, then the strategic
  connection game $\game$ has a strong equilibrium.
\end{theorem}
\begin{proof}
  We formulate the mixed integer program for the problem in directed
  graphs. It is simple to adjust it to undirected graphs, where we use
  only one variable $y_{ij}$ for each (undirected) edge $e = (i,j) \in
  E$.
  \begin{equation*}
    \renewcommand{\arraystretch}{1.7}
    \begin{array}{llll}
      \mbox{Min } & \multicolumn{3}{l}{\D \sum_{(i,j) \in E} c_{ij}y_{ij} } \\
      \mbox{s.t. } & \multicolumn{2}{l}{\D \sum_{\{j\;|\;(i,j) \in E\}} f_{ij}^k - \sum_{\{i \; | \; (i,j) \in E\}} f_{ji}^k \ge 1} & \mbox{ for $i = s_k$} \\
      & \multicolumn{2}{l}{ \D \sum_{\{j\;|\;(i,j) \in E\}} f_{ij}^k - \sum_{\{i \; | \; (i,j) \in E\}} f_{ji}^k \ge 0} & \mbox{ for $i \neq s_k, t_k$} \\
      & y_{ij} - f_{ij}^k \ge 0 & \multicolumn{2}{l}{\mbox{ for } (i,j) \in E, k \in K}\\
      & f_{ij}^k \ge 0, y_{ij} \in \{0,1\} & \multicolumn{2}{l}{\mbox{ for } (i,j) \in E, k \in K}
    \end{array}
  \end{equation*}
  In this MIP we optimize for each player $k$ a flow, which is
  required to have value 1 by the constraints at the source, and which
  can only exit through the sink. The individual flows are coordinated
  by capacity constraints $y_{ij} - f_{ij}^k \ge 0$. Each edge that is
  used by at least one player fractionally has to be fully paid for in
  the objective function. We can relax this program by using $y_{ij}
  \ge 0$. Then the dual can be formulated using variables $\delta_i^k$
  for the flow conservation constraints and $\gamma_{ij}^k$ for the
  coordination constraints. Intuitively, the values $\delta_i^k$
  introduce a node potential of contributions, and $\gamma_{ij}^k$ can
  be seen as contributions towards the edges that are bought.
  
  It has been observed in~\cite{Tamir91} that this is program is
  wihtin Owens linear production model. Hence, if the integrality gap
  is 1, the optimal dual solution yields a core solution. Using
  similar arguments as before, we can also show that in this case a SE
  exists. In particular, each player pays $s_k(i,j) =
  y_{ij}^*(\gamma_{ij}^{k*})$ towards edge $(i,j)$. For a coalition $C$
  we can again reduce costs of edges by removing players of $K-C$. Due
  to the additive structure of the LP, the primal and dual optimal
  solutions remain optimal for the reduced LPs. This means no
  coalition can reduce total payments, and no coalition can be
  violating. This proves the theorem.
\end{proof}

This insight allows us to derive one of the main results shown
in~\cite{Epstein09} in a simple and compact way.
\begin{theorem}{\cite{Epstein09}}
  For a single source connection game $\game$ on a directed
  series-parallel graph a strong equilibrium always exists and can be
  computed in polynomial time.
\end{theorem}
The existence result follows easily by observing that the Flow-LP for
single source games on directed series-parallel graphs has integrality
gap 1. A proof can be derived from~\cite{Prodon85}. Solving this LP
resembles the construction in~\cite{Epstein09}.

Another class, in which the above LP can be used to show existence of
SE, are games based on the minimum spanning tree (MST) problem. MST
games are single source games, in which in every vertex of $G$ is a
sink node for at least one player.

\begin{theorem}
  In every MST game $\game$ there is a strong equilibrium, which can
  be computed in polynomial time.
\end{theorem}

For the problem in directed graphs, a SE can be computed from dual
solutions of the LP~\cite{Tamir91}. One of these dual solutions is the
core solution derived for the original non-emptiness
proof~\cite{Granot81}. In this solution, each player $k$ pays exactly
for the unique arc of the tree leaving his sink $t_k$. This rule has
also been described by Bird~\cite{Bird76}. It requires an easy
argument to see that it yields a SE, even for the MST game in
undirected graphs.

While in these cases we have guaranteed existence and efficient
algorithms to compute SE, the problem of deciding the existence of SE
is \classNP-hard. This follows from a simple adjustment, which allows
to interpret UFL games as single source connection games on directed
graphs. 

\begin{corollary}
  It is \classNP-hard to decide if a given single source connection
  game $\game$ on a directed graph has a strong equilibrium.
\end{corollary}

\section{Strong Equilibria beyond Linear Programming}
\label{sec:Comb}
\subsection{Connection Games}

For set cover and facility location games the integrality gap
condition provides a complete characterization of games $\coopGame$
having core solutions. With our theorems we obtain a complete
characterization also for the existence of SE in strategic games
$\game$. For network design games like the connection game, the
integrality gap condition is sufficient to show existence of SE and
non-emptiness of the core, but it is not necessary. A tight
characterization of games with non-empty core has not been obtained so
far.

For strategic games and SE it has been shown in~\cite{Epstein09} that
there is a single source connection game without SE, but the
corresponding cooperative game to their example does not allow a core
solution as well. By Proposition~\ref{prop:SEisCore}, however, this is a
prerequisite for SE existence. Coalitional connection games with an
empty core (and thus without SE) have already been presented
in~\cite{Granot98}. We here show that even a spanning property of the
optimum solution $R^*$ is not sufficient to guarantee SE existence or
to obtain SE from core solutions. This implies that the relation
between core and SE is not as robust as for the other games considered
previously. A complete characterization of the existence of SE in
(single source) connection games remains as an open problem.

\begin{lemma}
  \label{lem:SteinerNo}
  There are corresponding strategic and coalitional single source
  connection games $\game$ and $\coopGame$ such that $R^*$ is a MST of
  $G$ and $\coopGame$ has a core solution but $\game$ has no strong
  equilibrium.
\end{lemma}

\begin{proof}
  Our example game is shown in Fig.~\ref{fig:noSNE}. It is based on a
  game presented in~\cite{Granot98}, which consisted only of the three
  lower layers up to node $s'$. It was shown that this game has an
  empty core, but $R^*$ passes through all vertices of $G$. This also
  implies that there can be no SE.
  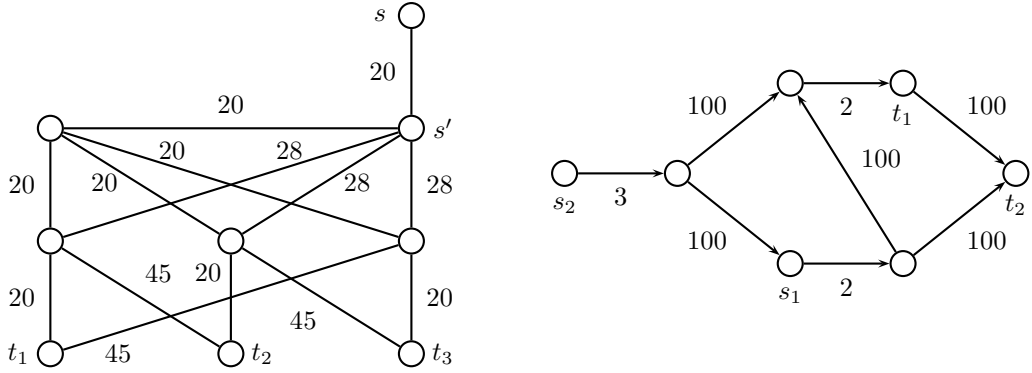
\begin{figure}
    \begin{center}
      \psset{unit=0.06cm}
      \begin{pspicture}(0,0)(107,85)
        \cnode(5,5){3}{A}
        \cnode(45,5){3}{B}
        \cnode(85,5){3}{C}

        \cnode(5,30){3}{D}
        \cnode(45,30){3}{E}
        \cnode(85,30){3}{F}

        \cnode(5,55){3}{G}
        \cnode(85,55){3}{H}

        \cnode(85,80){3}{I}

        \ncline{A}{D}
        \naput{20}
        \ncline{B}{E}
        \rput(40,23){20}
        \ncline{C}{F}
        \nbput{20}

        \ncline{A}{F}
        \rput(20,5){45}
        \ncline{B}{D}
        \nbput{45}
        \ncline{C}{E}
        \naput{45}

        \ncline{D}{G}
        \naput{20}
        \ncline{E}{G}
        \rput(17,43){20}
        \ncline{F}{G}
        \rput(32,50){20}

        \ncline{D}{H}
        \rput(58,50){28}
        \ncline{E}{H}
        \rput(73,43){28}
        \ncline{F}{H}
        \nbput{28}

        \ncline{G}{H}
        \naput{20}

        \ncline{H}{I}
        \naput{20}
        
        \rput(78,80){$s$}
        \rput(92,55){$s'$}
        \rput(-2,5){$t_1$}
        \rput(52,5){$t_2$}
        \rput(92,5){$t_3$}

      \end{pspicture}
      \begin{pspicture}(0,0)(107,85)
        \cnode(10,45){3}{A}
        \cnode(35,45){3}{B}
        \cnode(60,25){3}{C}
        \cnode(60,65){3}{D}
        \cnode(85,25){3}{E}
        \cnode(85,65){3}{F}
        \cnode(110,45){3}{G}

        \ncline{->}{A}{B}
        \nbput{3}
        \ncline{->}{C}{E}
        \nbput{2}
        \ncline{->}{D}{F}
        \nbput{2}

        \ncline{->}{B}{C}
        \nbput{100}
        \ncline{->}{B}{D}
        \naput{100}
        \ncline{->}{E}{D}
        \nbput{100}
        \ncline{->}{E}{G}
        \nbput{100}
        \ncline{->}{F}{G}
        \naput{100}

        \rput(10,38){$s_2$}
        \rput(110,38){$t_2$}
        \rput(60,18){$s_1$}
        \rput(85,58){$t_1$}

      \end{pspicture}

    \end{center}
    \caption{\label{fig:noSNE} Left: A single source connection game
      with 3 players, a non-empty core, but without a SE. $R^*$ is an
      MST of $G$ and consists of all edges of cost 20. Right: A
      multicut game on a directed graph with 2 players and a non-empty
      core. The game has no NE.}
  \end{figure}

  To obtain our game in Fig.~\ref{fig:noSNE}, we added the new source
  $s$ and an edge of cost 20 to the old source $s'$. Then the
  constraints for the contributions of the coalitions allow a feasible
  cost sharing by assigning each player a share of $160/3 \approx
  53.33$. This removes the incentives to deviate on a global scale,
  which is sufficient for non-emptiness of the core. On a local scale,
  however, the instable structure up to $s'$ is still intact. The
  additional contributions towards $(s',s)$ do not change the
  strategic incentives within the lower parts of the graph. It can be
  verified that in this game no SE exists. This proves the lemma.
\end{proof}

\subsection{Terminal Backup Games}

In this section we study games based on the terminal backup
problem~\cite{Anshe09,Anshe07}. In this game there is a graph
$G=(V,E)$, each player is a vertex ($K \subset V$), and resources are
the edges with costs $c(e) \ge 0$. Each player strives to be connected
to at least $d-1$ other player vertices, for $d \ge 2$. It has been
shown in~\cite{Anshe07} that the terminal backup problem can be solved
in polynomial time for $d=2$. Here we show that every core solution
can be turned into a SE for these games. In addition, we show how to
decide if a game has a SE and how to obtain SE in polynomial time if
they exist.

\begin{theorem}
  \label{theo:TB2}
  If a terminal backup game $\coopGame$ with $d = 2$ has a non-empty
  core, then the strategic game $\game$ has a strong equilibrium.
\end{theorem}

\begin{proof}
  Suppose there is a core solution in $\coopGame$, but there is no SE
  in $\game$. We can allocate the cost shares from the core for each
  player $k$ to a consecutive subtree starting at their
  vertex. Furthermore, we can adjust the graph such that $R^*$ is only
  composed path components with two player vertices at the ends, or of
  star components with at most three player vertices. This adjustment
  can be achieved by replacing each non-player vertex with a clique of
  sufficiently many vertices and clique edges of cost 0.

  In a star component, each player pays a cost of at most the
  connection to the center. For contradiction assume that the cost
  share of player $i$ in the core solution is $\epsilon > 0$ larger
  than the cost of the connection to the star center. Then every other
  player in the component must pay at most its' own connection cost to
  the star center minus $\epsilon$. Thus, either the component is a
  path, or player $i$ is left to pay at least $2\epsilon$, which is a
  contradiction. This shows that there exists an allocation of core
  cost shares to the edges such that each player pays for a
  consecutive path starting at his terminal.

  For the sake of contradiction assume that this allocation is not a
  SE. Suppose $C$ is a violating coalition of players. In their
  improvement the players of $C$ can improve by changing their
  connections and create a new component. If this new component is
  paid for fully by the players in $C$, this corresponds to a
  constraint considered for the core solution. Hence, the players in
  such a new component cannot all profit from such a deviation. 

  On the other hand, suppose players use edges to create their new
  component, for which part of the cost is paid for by players not in
  $C$. These other players pay a consecutive subtree starting at their
  terminals. Hence, their connection requirement will be satisfied in
  the new component as well. Thus, they and all the players located in
  the bought subtree incident to their vertex can be added to the
  coalition and the deviation. The costs can be redistributed such
  that everybody improves and pays a lesser amount than before. This
  again yields a set of improving players that pays completely for
  their component. However, as such deviations are covered by the core
  constraints, this is again a contradiction to the cost shares being
  a core solution. This completes the proof of the theorem.
\end{proof}

The above property allows us to efficiently determine if SE and core
solutions exist and to compute them in polynomial time if they exist.

\begin{corollary}
  There is a polynomial time algorithm to determine if a coalitional
  terminal backup game $\coopGame$ with $d=2$ has a core solution and
  if the strategic game $\game$ has a strong equilibrium. If they
  exist, a core solution and a strong equilibrium can be computed in
  polynomial time.
\end{corollary}

\begin{proof}
  We can compute an optimal solution in polynomial time. We then
  decide if a core solution exists as follows. The structure of the
  problem implies that optimal solutions can be assumed as
  compositions of components for 2 or 3 players. There are only a
  polynomial number of such coalitions, and the optimum solution for
  each such coalition can be found in polynomial time for each of
  them. Hence, the set of inequalities that characterizes the core is
  only of polynomial size and can be obtained in polynomial
  time. Thus, we can check in polynomial time if this set of
  inequalities has a solution and in this way obtain a member of the
  core. Given a core solution, we can use the computed optimum
  solution and our structural insight about SE to find the appropriate
  allocation of payments to edges in polynomial time.
\end{proof}

For larger connectivity requirements of $d \ge 4$ we construct games
where the consecutive payment condition of Theorem~\ref{theo:TB2} is
violated. In this case, a core solution cannot be turned into a SE.

\begin{lemma}
  For any $d \ge 4$ there is a coalitional terminal backup game
  $\coopGame$ with a core solution and a corresponding strategic game
  $\game$ without a strong equilibrium.
\end{lemma}

In fact, our example game can be derived directly with the
single source connection game in Fig.~\ref{fig:noSNE} above. We simply
replace the source $s$ by a clique of 4 or more terminals and 0-cost
edges.

\subsection{Network Cutting Games}

In this section we briefly discuss a network cutting game, in which
there is a graph $G=(V,E)$ and each player strives to disconnect a
subset $S_k \subset V$ from another subset $T_k \subset V$. Each edge
$e \in E$ has a cost $c(e) > 0$ for disconnection. This approach
yields coalitional and strategic games based on a variety of
minimum-cut problems like $s$-$t$-cut, multicut, mulitway cut, etc. It
is introduced and studied with respect to NE in the special cases of
mulitway cut and multicut in~\cite{AnshelevichCUT10}.

More formally, for each player $k$ denote by $\mathcal{P}_k$ the set
of all paths in $G$ from a node in $S_i$ to a node in $T_i$. When we
introduce a variable $x_e$ for each edge $e \in E$, then for each path
$P \in \mathcal{P}_k$ player $k$ has the constraint $\sum_{e \in P}
x_e \ge 1$. Note that these are simple 0/1-covering constraints, and
thus the resulting integer program is a special case of the set cover
integer program presented above. In particular, we can simply regard
paths as elements and edges as sets. This implies that if the
integrality gap is 1, we have existence of core solutions and SE. For
instance, this holds on directed and undirected graphs for
single-source games that have $S_i = \{s\}$ for each $i \in K$.

\begin{theorem}
  If the Covering-LP has an integral optimum solution, then the
  strategic network cutting game $\game$ has a strong equilibrium.
\end{theorem}

Note that there is a subtle twist to this observation. While in the
set cover game every element (i.e., every path) is a player, in the
cutting game players strive to cover multiple elements (i.e., cut
multiple paths). The previous theorem still holds, because by
clustering elements we simply reduce the granularity of possible
coalitions to those, which can be obtained by the union of sets
$\mathcal{P}_k$. In fact, by this transformation we increase the set
of games that allow a strong equilibrium and a core solution.
\begin{proposition}
  There are network cutting games $\game$ with strong equilibria, for
  which the underlying network cutting problem has an integrality gap
  of more than 1.
\end{proposition}
\begin{proof}
  Consider a network multiway cut game, in which every player $i$ has
  a vertex $s_i \in V$ and wants to disconnect it from every other
  player vertex, i.e., $T_i = \{s_j : i \neq j \in K \}$. Consider a
  star, in which the player vertices are exactly the leaves and all
  edges have cost 1. This class of instances is known to have the
  maximum integrality gap of $2-1/k$ for the covering LP of the
  network multicut problem. In particular, the fractional optimum
  solution assigns each edge to be in the cut with $x_e = 1/2$, while
  the integral optimum fully cuts all but one edge. In a SE we pick
  one player $i$ to be \emph{uncut}. Each other player $j \neq i$ is
  assigned to purchase the edge incident to $s_j$ completely. Note
  that every coalition $C$ without the uncut player $i$ must pay at
  least $k$ to remain disconnected from $i$. Every other coalition
  must pay at least $k-1$. Hence, no coalition can reduce their
  payments in sum, and the existence of a SE and a core solution
  follows.
\end{proof}
A similar observation can be made for multiway cut games, in which
geometric LP relaxations~\cite{Calinescu00} have an integrality gap of
more than 1. In fact, it can be observed that the arguments
from~\cite{AnshelevichCUT10} for existence of optimal NE can be
extended in a straightforward way to show that every mutiway cut game
on an undirected graph admits a SE. In general network cutting games,
however, the set of strategic games with a SE is not equivalent to the
set of cooperative games with a non-empty core.

\begin{lemma}
  There are corresponding coalitional and strategic network cutting
  games $\coopGame$ and $\game$ such that $\coopGame$ has a core
  solution but $\game$ has no strong equilibrium.
\end{lemma}

\begin{proof}
  For undirected graphs we consider two players and a star graph. We
  set $S_1 = \{s_1\}$, $S_2 = \{s_2\}$, $T_1 = \{t_1\}$ and $T_2 =
  \{s_1,t_1\}$. The edge costs to the center node $u$ are $c(s_1,u) =
  c(t_1,u) = 2$ and $c(s_2,u) = 3$. The set of core solutions is
  $\gamma_1 = 2-\epsilon$ and $\gamma_2 = 2 + \epsilon$ for $0 \le
  \epsilon \le 1$. Note that the unique optimum solution is to cut
  $(s_1,u)$ and $(t_1,u)$. In such a solution, however, if $|s_1| >
  0$, player 1 can unilaterally improve by removing the larger of his
  payments. Player 2 does not pay for both edges, because paying only
  for $(s_1,u)$ is cheaper.

  For directed graphs we can even leave $T_2 = \{t_2\}$ as a
  singleton. We transform the graph to the one shown in
  Fig.~\ref{fig:noSNE}. A similar argument shows non-existence of SE.
\end{proof}

This construction implies that when we relax the assumption that
\emph{every} element or terminal is a player in a set cover or
facility location game, equivalence between core and SE does not hold
anymore. On another note, the proof shows absence of NE in general
strategic network cutting games on undirected games. For directed
graphs the absence of NE holds true even for minimum multicut games,
in which $S_i$ and $T_i$ are singleton sets for all players $i \in K$.

\section{Extensions}

\subsection{Fractional and Non-Binary Resources}
\label{sec:Fractional}

Apart from equivalence of core and SE, a natural question is to
characterize cases when we can derive SE from core solutions using
linear programming. This was possible when the integrality gap was 1
in all the cases described above. With exception of the CCRFL games,
all games studied above yield linear constraints that fall into one of
two classes. One type of constraint is $\sum_i x_i \ge 1$, i.e., a
simple covering constraint with 0/1 coefficients, by which we can
express exactly the vertex and set cover games. The other constraint
type is $y_i - \sum_j x_{ij} \ge 0$, i.e., a coordination constraint
that requires a resource to become bought when at least one player
uses it. This second type of constraint allows us to treat facility
location and network design cut games. What happens if we slightly
generalize these constraints?

As an example let us first consider dropping the integrality
requirement. Using results on Owen's linear production model one can
show, for instance, that vertex cover games always allow a core
solution if vertices and sets can be bought in \emph{fractional}
amounts. Does a SE also exist for strategic games in these cases? To
answer this question we must adjust the strategic game to allow
vertices to be bought fractionally. The obvious adjustment is to
assign a fraction proportional to the total payment. In a state $s$ of
the strategic \emph{fractional vertex cover game} a vertex $v$ is
bought to the degree $x_v = \sum_{k \in K} s_k(e) / c(v)$. For a
player $k$ corresponding to edge $e = (u,v)$ the individual cost is
$|s_k|$ if $x_u + x_v \ge 1$ and prohibitively large otherwise.

A second, closely related variant is the case when we keep the
integrality condition, but we increase the covering requirements and
allow multiple units of a resouce to be bought. In particular, we
change the constraints to a type $\sum_i x_i \ge b$, where $b > 0$ and
$x_i \in \nat$. As for the fractional games the total payments of the
players determine the number of units bought of a resource. We term
these games \emph{non-binary vertex cover games}. More formally, in a
state $s$ we have $x_u = \lfloor \sum_{k \in K} s_k(u)
\rfloor$. Player $k$ corresponding to edge $(u,v)$ has a required
coverage of $b_k \in \nat$ and individual cost $c_k(s) = |s_k|$ if
$x_u + x_v \ge b_k$ and prohibitively large otherwise.

Note that for both of these game classes
Proposition~\ref{prop:SEisCore} and Theorem~\ref{theo:SPoA} continue
to hold. In contrast to our results above, however, we show next that
there might be no SE -- although non-emptiness of the core can be
established via the same linear programming machinery that was used
before.

\begin{theorem}
  There are corresponding strategic and coalitional fractional or
  non-binary vertex cover games $\coopGame$ and $\game$ such that
  $\coopGame$ has a core solution but $\game$ has no strong
  equilibrium.
\end{theorem}

\begin{proof}
  For both variants the proof follows with a triangle, vertex costs
  $c(u) = 3$, $c(v) = 5$, and $c(w) = 7$, and players 1 to 3
  corresponding to edges $(u,w)$, $(u,v)$ and $(v,w)$,
  respectively. 

  In the fractional game the unique optimum solution to the underlying
  vertex cover problem is $x_u^* = x_v^* = x_w^* = 1/2$, and the
  unique core solution is $\gamma_1 = 2.5$, $\gamma_2 = 0.5$ and
  $\gamma_3 = 4.5$. Theorem~\ref{theo:SPoA} yields that $x^*$ has to
  be purchased in every SE, but no player is willing to contribute to
  $w$. We obviously must have $s_2(w) = 0$. If $s_1(w) > 0$, player 1
  can deviate unilaterally and achieve the amount $s_1(w) / 7$ of
  coverage by contribution to $u$ with less payments. The same holds
  for player 2 and vertex $v$.

  For the non-binary version, we set all covering requirements to $b_1
  = b_2 = b_3 = 4$. Then the unique optimum $x^*$ to the underlying
  vertex cover problem and the unique core payments $\gamma$ are the
  same as before scaled by factor 4. Observe that we have an
  integrality gap of 1 in this game. The core solution is unique, so
  with Proposition~\ref{prop:SEisCore} we know that in every SE $|s_1|
  = 10$ and $|s_2| = 2$. This implies $4 \le s_1(w) \le 6$. By
  removing this payment from $w$, player 1 reduces the number of units
  bought of $w$ by exactly 1. However, he can obtain an additional
  unit of $u$ at a cost of 3. This yields a profitable unilateral
  deviation and proves the theorem.
\end{proof}
  
This shows that in the class of non-binary vertex cover games neither
non-emptiness of the core nor an integrality gap of 1 can guarantee
the existence of SE.

\subsection{Approximate Equilibria}
\label{sec:Approx}

We have presented a method to derive SE in strategic cost sharing
games via linear programming. A disadvantage of the concept of SE is
that they might not exist in a game. However, our approach proves to
be applicable even to approximate SE. Using primal-dual algorithms we
can compute $(\alpha,\beta)$-SE with small (constant) ratios in
polynomial time for vertex cover, set cover, and facility location
games. The proof for the following theorem can be derived directly
from arguments in~\cite{HoeferISAAC06,CarHoe06}. 
\begin{theorem}
  There are efficient primal-dual algorithms to compute $(2,2)$-SE for
  vertex cover, $(f,f)$-SE for set cover (where $f$ is the maximum
  frequency of any element in the sets), and $(3,3)$-SE for metric UFL
  games in polynomial time.
\end{theorem}
We provide a sketch of the proof
here. In~\cite{HoeferISAAC06,CarHoe06} we have observed that the
results stated in the theorem hold for $(\alpha,\beta)$-NE with the
same ratios. Consider the proof for, e.g., the (2,2)-NE in vertex
cover games in~\cite{CarHoe06}. The primal-dual algorithm iteratively
raises payments made by single edges until one of the incident
vertices becomes bought. Players can own multiple edges, but their
total payments are made up by the sum of payments assigned to their
edges. In fact, the payments assigned by the algorithm are independent
of which player owns which edge. A deviation of a player owning
multiple edges is equivalent to a coordinated deviation the coalition
of single edge players. Thus, the proof that the algorithm computes
(2,2)-NE shows that the state computed by the algorithm allows no
coalitional deviation of single edge players that decreases their
\emph{payments in sum} by a factor of strictly more than 2. In this
case we obviously cannot have a deviation in which \emph{every} player
of a coalition reduces his cost by a factor of strictly more than
2. These observations yield the result for vertex cover games, and
similar arguments can be applied using the algorithms for set cover
and facility location games treated in~\cite{HoeferISAAC06}.

\section{Conclusions and Open Problems}
\label{sec:Conclude}

In (single source) connection, network cutting and fractional and
non-binary games the use of linear programming duality does not
necessarily yield a complete characterization of the games that admit
SE. In these games and other interesting variants of cost sharing in
network design our work opens up numerous interesting research
problems regarding the characterization and computation of exact and
approximate SE.

More generally, we believe that the linkage between core and strong
equilibrium could be present in other cost sharing games, which go
beyond the classes of games treated in this paper. Exploring these
classes of games is an interesting avenue for further research. More
concretely, our games have linear programming formulations that lie
within Owens linear production model. Non-emptiness of the core,
however, can also be shown within a more general class of
problems. This more general framework, termed generalized linear
production model in~\cite{Granot86}, has a non-additive structure, and
it encompasses for instance the cut-based LP-formulation for Steiner
Network problems~\cite{Skorin95}. It is a fascinating open problem to
see if this framework can also be used to derive exact and approximate
SE in strategic cost sharing games.

\subsubsection*{Acknowledgement}
The author would like to thank Elliot Anshelevich and Bugra Caskurlu
for valuable discussions and feedback about the results in this paper.

\bibliography{../../../Bibfiles/literature,../../../Bibfiles/martin}
\bibliographystyle{plain}

\appendix
\section{Connection-Restricted Facility Location}

In a \emph{CRFL problem} there is a set $T$ of terminals and a set $F$
of facilities. We set $n_t = |T|$ and $n_f = |F|$. In addition to the
UFL problem each facility has a set of allowable subsets
$\mathcal{A}_f \subseteq 2^T$. The goal is to open a subset of
facilities and buy a set of connections of minimum total cost, such
that each terminal is connected to an opened facility, and the set of
terminals connected to each opened facility $f$ is in
$\mathcal{A}_f$. In the \emph{CRFL game} each player owns a terminal,
i.e., $K = T$. The constraint of player $t$ is satisfied if there is a
bought connection $(t,f)$ to some opened facility $f$, and the subset
of terminals that have a bought connections to $f$ is in
$\mathcal{A}_f$. We can formalize the CRFL problem by an integer
program as follows:
\begin{equation*}
  \begin{array}{lll}
    \mbox{Min } & \D \sum_{f \in F} c(f)y_f + \sum_{t \in T} c(t,f)x_{tf} \\
    \mbox{subject to } & \D \sum_{f \in F} x_{tf} \ge 1 & \mbox{ for all }
    t \in T\\
    & (y_f, x_{1f}, \ldots, x_{n_tf}) \in \mathcal{A}_f & \mbox{ for
      all } f \in F\\  
    & y_f, x_{tf} \in \{0,1\} & \mbox{ for all } t \in T, f \in F, 
  \end{array}
\end{equation*}
where\\ $\mathcal{A}_f = \{(0,\ldots,0)\} \cup \{ (1,\chi_{A_f})\;|\;
A_f \subseteq T \mbox{ feasible for } f\} \subseteq \{0,1\}^{n_t+1}$,
and $\chi_{A_f}$ denotes the characteristic vector of the subset
$A_f$.

We here concentrate on a subclass of \emph{closed} games (denoted
CCRFL). In these games the sets $\mathcal{A}_f$ are downward closed,
i.e., for every $A \subseteq A' \in \mathcal{A}_f$ we have $A \in
\mathcal{A}_f$. Note that this class encompasses a large variety of
facility location problems considered in the literature, e.g., with
capacity or incompatibility constraints.

\begin{theorem}
  If a CCRFL game $\coopGame$ has a non-empty core, then the strategic
  game $\game$ has a strong equilibrium.
\end{theorem}

\begin{proof}
  Following the argumentation in~\cite{GoeSku04} it is possible to use
  the conic hull of the sets $\mathcal{A}_f$ to derive a linear
  relaxation:
  \begin{equation*}
    \begin{array}{lll}
      \mbox{Min } & \D \sum_{f \in F} c(f)y_f + \sum_{t \in T} c(t,f)x_{tf} \\
      \mbox{subject to } & \D \sum_{f \in F} x_{tf} \ge 1 & \mbox { for all }
      t \in T\\
      & (y_f, x_{1f}, \ldots, x_{n_tf}) \in \mbox{cone}(\mathcal{A}_f) &
      \mbox{ for all } f \in F. 
    \end{array}
  \end{equation*}
  For this program a dual can be given by
  \begin{equation*}
    \begin{array}{ll}
      \mbox{Max } & \D \sum_{t \in T} \gamma_t \\
      \mbox{subject to } & \D \sum_{t \in A_f} \gamma_t \le c(f) + \sum_{t
        \in A_f} c(t,f) \\ 
      & \mbox{ for } f \in F \mbox{ and $A_f \in \mathcal{A}_f$.}
    \end{array}
  \end{equation*}
  Now we can apply similar arguments as before. An integral optimum
  solution $(x^*, y^*)$ to the LP-relaxation represents a partition of
  the terminal set $T$ into a collection of feasible sets $A_f^*$, one
  for each facility $f$. The constraints corresponding to these sets
  hold with tightness, and we can assign each player $t$ to pay for
  her terminal the amount $s_t(t,f) = \min\{\gamma_t^*,c(t,f)\}$ as
  connection cost to $f$ with $t$ connected to $f$, in which
  $\gamma^*$ is the optimum solution to the dual. For the opening
  costs $s_t(f) = \max\{\min\{c(f), \gamma_t^* - c(t,f)\}, 0\}$. Note
  that such an assignment is always possible due to $\mathcal{A}_f$
  being downward closed. In particular, no player $t$ is required to
  pay for the connection cost of any other player. Thus, no coalition
  of players can improve by simply dropping payments.

  In total this pays exactly for all costs of the solution by
  duality. Suppose there is a violating coalition $C$. This coalition
  must be able to connect their terminals differently at a cheaper
  total cost. Consider the strategy vector after the coalition has
  changed its strategy. Each member $t' \in C$ must again be part of
  some $A_{f'}$ for some facility $f'$, for which the total
  (connection + opening) costs are fully paid for. In particular, the
  new payments exactly pay for $c(f') + \sum_{t \in A_{f'}}
  c(t,f')$. Note that no player has increased his payments, but $t'$
  has strictly decreased his payments. This means that the original
  payments coming from $\gamma_t^*$ violate the dual constraint
  corresponding to $\mathcal{A}_{f'}$. This is a contradiction to
  $\gamma^*$ being the optimal dual solution.
\end{proof}

The recognition of SE can be done similarly as for UFL games.

\begin{corollary}
  Given a strategy vector for a CCRFL game $\game$ we can verify in
  polynomial time if it is a SE.
\end{corollary}

\end{document}